\documentclass[preprint,5p,times,twocolumn,10pt]{elsarticle}

\usepackage{amssymb}

\journal{Physics Letters A}

\begin{document}

\newcommand{\beq}{\begin{equation}}
\newcommand{\eeq}{\end{equation}}
\newcommand{\beqa}{\begin{eqnarray}}
\newcommand{\eeqa}{\end{eqnarray}}
\def\ket#1{|\,#1\,\rangle}
\def\bra#1{\langle\, #1\,|}
\def\braket#1#2{\langle\, #1\,|\,#2\,\rangle}
\def\proj#1#2{\ket{#1}\bra{#2}}
\def\expect#1{\langle\, #1\, \rangle}
\def\trialexpect#1{\expect#1_{\rm trial}}
\def\ensemblexpect#1{\expect#1_{\rm ensemble}}
\def\kpsi{\ket{\psi}}
\def\kphi{\ket{\phi}}
\def\bpsi{\bra{\psi}}
\def\bphi{\bra{\phi}}

\begin{frontmatter}

\title{Local transformations of superpositions of entangled states}

\author{Iulia Ghiu}

\address{Centre for Advanced Quantum Physics, University of Bucharest, PO Box MG-11,
R-077125, Bucharest-M\u{a}gurele, Romania}

\begin{abstract}
Suppose that we have two entangled states $\ket
{\phi_1}$, $\ket{\psi_1}$ that cannot be converted to any of other two states $\ket{\phi_2}$, $\ket{\psi_2}$ by local operations and classical communication. We analyze the possibility of locally transforming a superposition of $\ket{\phi_1}$ and $\ket{\psi_1}$ into a superposition of
$\ket{\phi_2}$ and $\ket{\psi_2}$. By using the Nielsen's theorem we find the necessary and sufficient conditions for this conversion to be performed.
\end{abstract}

\begin{keyword}
entanglement transformations \sep superposition
\PACS 03.65.Ud \sep 03.67.-a
\end{keyword}

\end{frontmatter}

\section{Introduction}

The entangled states are the main resources in many processes of quantum processing, such as quantum cryptography, quantum teleportation, quantum telecloning, superdense coding or quantum computation \cite{Nielsenbook}. In order to perform some tasks it is useful to manipulate the entanglement under special conditions, namely allowing only local operations and classical communication (LOCC). The method of finding the possible transformations between bipartite entangled states by LOCC was found by Nielsen \cite{Nielsen} and is based on the theory of majorization.
Let $\ket{\Psi}=\sum_{j=1}^d\sqrt{\alpha_j}\ket{jj}$ and $\ket{\Phi}=\sum_{j=1}^d\sqrt{\beta_j}\ket{jj}$ be two bipartite states whose Schmidt coefficients are ordered in decreasing order: $x_1\ge x_2\ge ... x_d$ ($x=\alpha$ and $\beta$).
Then $\ket{\Psi} \to \ket{\Phi }$ by LOCC if and only if $\alpha $ is majorized by $\beta $, $\alpha \prec \beta $, i.e. if for each $k=1,...,d$ \cite{Nielsen}
\beq
\sum_{j=1}^k\alpha_j \le \sum_{j=1}^k\beta_j.
\label{maj}
\eeq
The Nielsen's theorem for entanglement manipulation has many applications in the process of the catalysis, asymptotic entanglement transformations or for the local distinguishability of states.

A major interest in entanglement transformations has been the catalysis. This enables the conversion between two initially inconvertible entangled states assisted by a lent entangled state, which is recovered at the end of the process \cite{Plenio}. The Nielsen's theorem was recently applied for finding the conditions for a state to be a general catalyst by Song {\it et al.} \cite{Song}. In a previous work of us \cite{Ghiu}, we proved that bipartite and tripartite states cannot be used as catalysis states to enable local transformations between the entangled states which belong to the two inequivalent classes of three-particle states: the GHZ class and the W class. The inequivalent classes of entangled states have recently been investigated with a renewed interest: Chattopadhyay and Sarkar have shown that there is an infinite number of pure entangled states with the same entanglement, but all being incomparable to each other (i.e. being members of inequivalent classes) \cite{Ch}. Asymptotic entanglement manipulations were analysed by Bowen and Datta, who considered different measures for finding bounds on optimal rates of local entanglement conversions \cite{Datta}.

The local distinguishability of orthogonal states was studied by Horodecki {\it et al.} by proposing a method that involves pure states \cite{Horodecki}. The key element of their method is the building of a state in a larger system with the help of a {\it superposition} instead of a mixture, which was previously used in the scientific literature. The local indistinguishability is proved by showing that a transformation is impossible under LOCC due to the Nielsen's theorem. Recently Fan found a general approach for distinguishing arbitrary bipartite states, i.e. not only entangled states but also separable ones, by LOCC \cite{Fan}.

Another fundamental problem in quantum information theory is the relation between the entanglement of a given state and the entanglement of its individual terms. The entanglement of superposition of states was investigated by Linden {\it et al.}, who found upper bounds on the entanglement \cite{Popescu}.  The von Neumann entropy $E(\psi )=S(\mbox{Tr}_A\proj{\psi}{\psi})$ was employed as a measure of entanglement in this analysis.
Suppose that we have a superposition of two states:
$
\ket{\Gamma}=\alpha \ket{\phi}+\beta \ket{\psi}.
$
In the particular case when $\ket{\phi} $ and $\ket{\psi}$ are bi-orthogonal states, the following equality holds:
\beq
E(\Gamma )=|\alpha |^2E(\phi )+|\beta |^2E(\psi)|+h_2(|\alpha |^2),
\label{equality}
\eeq
where $h_2(x):=-x\log _2x-(1-x)\log _2(1-x)$ is the binary entropy function.
Many generalizations of this paper were recently given: lower and upper bounds on the entanglement of superposition \cite{Gour}, the entanglement measure used is the concurrence \cite{Cerf,Fan2}, the geometric measure and q-squashed entanglement in Ref. \cite{Song2}, multipartite entanglement \cite{Yu,Acin}, superpositions with more than two components \cite{Xiang}.
Despite of these generalizations, there are still many unsolved aspects regarding the entanglement of superpositions, one of them being the behavior under local manipulations.

In this Letter we analyze the following scenario: we start with two entangled states  $\ket{\phi_1}$, $\ket{\psi_1}$, which are inconvertible to any of other two entangled states $\ket{\phi_2}$, $\ket{\psi_2}$ by LOCC. We want to investigate the possibility of building a superposition of $\ket{\phi_1}$ and $\ket{\psi_1}$ that can be locally transformed to a superposition of $\ket{\phi_2}$ and $\ket{\psi_2}$.
It turns out that the necessary and sufficient conditions to make this conversion realizable involve some inequalities which have to be satisfied by the Schmidt coefficients of $\ket{\phi_1}$, $\ket{\psi_1},$ and $\ket{\phi_2}$.

The Letter is organized as follows. In section 2, we derive the main result, which consists of three propositions that represent the conditions for the local transformation of superpositions of entangled states. One example is given in subsection 2.2 to illustrate the application of our propositions. Finally, the conclusions are drawn in section 3.

\section{Transformation of superpositions of entangled states}

Suppose that we have two bipartite entangled states $\ket{\phi_1}$ and $\ket{\psi_1}$ with the Schmidt number equal to 2, such that they are bi-orthogonal. Consider other two bipartite, bi-orthogonal  entangled states with the Schmidt number equal to 2, $\ket{\phi_2}$ and $\ket{\psi_2}$, such that the first group of states cannot be transformed to any of the two states of the second group by LOCC:
\beqa
\ket{\phi_1}&\not\to &\ket{\phi_2}\nonumber\\
\ket{\phi_1}&\not\to &\ket{\psi_2}\nonumber\\
\ket{\psi_1}&\not\to &\ket{\psi_2}\nonumber\\
\ket{\psi_1}&\not\to &\ket{\phi_2}.
\label{notpos}
\eeqa
Let us define the two superpositions:
\beqa
\ket{\Gamma_1}&=&\sqrt{\alpha_1}\ket{\phi _1}+\sqrt{1-\alpha_1}\ket{\psi_1}; \nonumber\\
\ket{\Gamma_2}&=&\sqrt{\alpha_2}\ket{\phi _2}+\sqrt{1-\alpha_2}\ket{\psi_2}.
\label{superpos}
\eeqa
We address the following question: is there $\alpha_1$ and $\alpha_2$  such that the transformation $\ket{\Gamma_1}\to \ket{\Gamma_2}$ can be performed for {\it arbitrary} $\ket{\phi_j}$, $\ket{\psi_j}$, $j=1,2$? And if the transformation is possible,
what conditions should the coefficients $\alpha_1 $ and $\alpha_2$ satisfy?

It is well known that the entanglement cannot be increased by LOCC; this means that if $\ket{\Psi}\to \ket{\Phi}$, then $E(\Psi)\ge E(\Phi)$. Chattopadhyay {\it et al.} have recently proved that the entanglement of two comparable states $d\times d$ with $d\ge 3$ is different (Theorem 2 in \cite{Ch}), i.e. $E(\Psi)> E(\Phi)$.
Accordingly, by employing the equality (\ref{equality}), the necessary condition for performing the transformation $\ket{\Gamma_1}\to \ket{\Gamma_2}$ reads:
\beqa
&&h_2(\alpha_2)+\alpha_2\left[ E(\phi_2)-E(\psi_2)\right] +E(\psi_2)<h_2(\alpha_1)\nonumber\\
&&+\alpha_1\left[ E(\phi_1)-E(\psi_1)\right] +E(\psi_1),
\label{import}
\eeqa
where the entanglement is given by the von Neumann entropy.

Let us analyze the following example.
Suppose that the two initial bi-orthogonal entangled states are:
\beqa
\ket{\phi_1}&=&\sqrt{\frac{9}{10}}\; \ket{00}+\sqrt{\frac{1}{10}}\; \ket{11};\nonumber\\
\ket{\psi_1}&=&\sqrt{\frac{4}{5}}\; \ket{22}+\sqrt{\frac{1}{5}}\; \ket{33}.
\label{initial-ex}
\eeqa
The two final bi-orthogonal states are:
\beqa
\ket{\phi_2}&=&\sqrt{\frac{7}{10}}\; \ket{00}+\sqrt{\frac{3}{10}}\; \ket{11};\nonumber\\
\ket{\psi_2}&=&\sqrt{\frac{3}{5}}\; \ket{22}+\sqrt{\frac{2}{5}}\; \ket{33}.
\label{final-ex}
\eeqa
We can easily check by using the Nielsen's theorem that the conditions (\ref{notpos}) are fulfilled and that the entanglement of the four states is:
\beqa
E(\phi_1)&=&0.4690; \nonumber\\
E(\psi_1)&=&0.72192;\nonumber\\
E(\phi_2)&=&0.88129; \nonumber\\
E(\psi_2)&=&0.97095.
\eeqa

\begin{figure}
\includegraphics{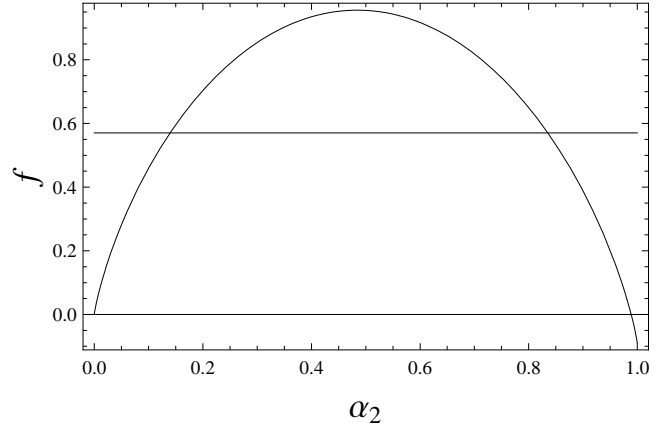}
\caption{Illustration of the inequality (\ref{inegalit-alfa2}): $f(\alpha_2)<0.57017$. We plot the function $f(\alpha_2)=h_2(\alpha_2)-0.08966\, \alpha_2$ and the constant function 0.57017. The solution of the inequality is $\alpha_2\in (0,0.1394)$ and $\alpha_2\in (0.8354,1)$ and this is the condition for $E(\Gamma_1)>E(\Gamma_2)$.}
\label{ineg}
\end{figure}

Let us choose $\alpha_1=\frac{3}{5}$. We have to determine $\alpha_2$ such that the inequality (\ref{import}) is verified:
\beq
f(\alpha_2)<0.57017,
\label{inegalit-alfa2}
\eeq
where $f(\alpha_2)=h_2(\alpha_2)-0.08966\, \alpha_2$.
The solution of this inequality is $\alpha_2\in (0,0.1394)$ and $\alpha_2\in (0.8354,1)$ (see Figure \ref{ineg})  and this is the condition for $E(\Gamma_1)>E(\Gamma_2)$.

Let us take $\alpha_2=0.85$.
One can compute the Schmidt coefficients of $\ket{\Gamma_1}$ and $\ket{\Gamma_2}$, respectively, and these are (we write them in a decreasing order):
\beqa
\ket{\Gamma_1}:&&\hspace{0.5cm}\frac{108}{200}; \hspace{0.3cm}\frac{64}{200}; \hspace{0.3cm}\frac{16}{200}; \hspace{0.3cm} \frac{12}{200};\nonumber\\
\ket{\Gamma_2}:&&\hspace{0.5cm}\frac{119}{200}; \hspace{0.3cm}\frac{51}{200}; \hspace{0.3cm}\frac{18}{200}; \hspace{0.3cm} \frac{12}{200}.
\eeqa
Since these Schmidt coefficients do not satisfy the majorization inequalities (\ref{maj}), it means that the transformation cannot be performed by LOCC.

We know that the decreasing of entanglement $E(\Gamma_1)>E(\Gamma_2)$ is not equivalent with the possibility of conversion between bipartite entangled states \cite{Nielsen}.
This example shows that there are $\alpha_1$ and $\alpha_2$, which satisfy the inequality (\ref{import}), i.e. $E(\Gamma_1)>E(\Gamma_2)$, and at the same time the conversion between these two states is not realizable by LOCC.

\subsection{The main result}

The bi-orthogonal states $\ket{\phi_j}$ and $\ket{\psi_j}$ with the Schmidt number equal to 2 have the following general expressions:
\beqa
\ket{\phi_1}=\sqrt{\xi_1}\ket{00}+\sqrt{1-\xi_1}\ket{11}, \hspace{1cm} \xi_1>\frac{1}{2};\nonumber\\
\ket{\psi_1}=\sqrt{\eta_1}\ket{22}+\sqrt{1-\eta_1}\ket{33}, \hspace{1cm} \eta_1>\frac{1}{2};\nonumber\\
\ket{\phi_2}=\sqrt{\xi_2}\ket{00}+\sqrt{1-\xi_2}\ket{11}, \hspace{1cm} \xi_2>\frac{1}{2};\nonumber\\
\ket{\psi_2}=\sqrt{\eta_2}\ket{22}+\sqrt{1-\eta_2}\ket{33}, \hspace{1cm} \eta_2>\frac{1}{2}.
\label{starile}
\eeqa
In addition we assume that
\beq
\xi_j>\eta_j,
\label{xieta}
\eeq
$j=1,2$. This is not a restriction, since if (\ref{xieta}) is not satisfied, we can convert $\ket{0}\leftrightarrow \ket{2}$ and $\ket{1}\leftrightarrow \ket{3}$ by LOCC and obtain the condition (\ref{xieta}). Also we impose that the conditions (\ref{notpos}) are verified and, due to the Nielsen's theorem, these are equivalent to $\eta_1>\xi_2$. Accordingly, due to the inequalities (\ref{xieta}), we get:
\beq
\frac{1}{2}<\eta_2<\xi_2<\eta_1<\xi_1<1.
\label{inetaxi}
\eeq

We show that the conversion $\ket{\Gamma_1}\to \ket{\Gamma_2}$ is possible for arbitrary $\ket{\phi_j}, \ket{\psi_j}$ by proving the following:
\begin{itemize}
\item{if $\xi_2 \in \bigg[ \frac{\xi_1\eta_1}{1-\xi_1+\eta_1},1\bigg)$, then $\alpha_1\in \bigg[ \frac{\eta_1}{1-\xi_1+\eta_1},\frac{\xi_2}{\xi_1}\bigg]$ ({\it Proposition 1});}
 \item{if $\xi_2 \in \bigg[ \frac{1-\eta_1}{2-\xi_1-\eta_1},\frac{\xi_1\eta_1}{1-\xi_1+\eta_1}\bigg)$, then $\alpha_1\in \bigg[ \xi_2,\frac{\eta_1}{1-\xi_1+\eta_1}\bigg)$ ({\it Proposition 2});}
 \item{if $\xi_2 \in \bigg( \frac{1}{2},\frac{1-\eta_1}{2-\xi_1-\eta_1}\bigg)$, then $\alpha_1\in \bigg[ \xi_2, \frac{1-\eta_1}{2-\xi_1-\eta_1}\bigg]$ ({\it Proposition 3}).}
\end{itemize}
In addition we must have
$
\alpha_2\ge \alpha_1\; \frac{\xi_1}{\xi_2},
$
which means that the first majorization inequality is the necessary and sufficient condition to enable the transformation.

{\bf Proposition 1.}

{\it Let $\ket{\phi_1}, \ket{\psi_1}, \ket{\phi_2}, \ket{\psi_2}$ be the states given by (\ref{starile}), being characterized by $\xi_j, \eta_j$ satisfying (\ref{inetaxi}). If $\xi_2 \in \bigg[ \frac{\xi_1\eta_1}{1-\xi_1+\eta_1},1\bigg)$, then for
$
\alpha_1\in \bigg[ \frac{\eta_1}{1-\xi_1+\eta_1},\frac{\xi_2}{\xi_1}\bigg]
$
and $\alpha_2>\frac{1}{2}$ we have the following:}
\beqa
&&\sqrt{\alpha_1}\ket{\phi_1}+\sqrt{1-\alpha_1}\ket{\psi_1}\to \sqrt{\alpha_2}\ket{\phi_2}+\sqrt{1-\alpha_2}\ket{\psi_2}\nonumber \\
&&by \hspace{0.2cm} LOCC
\hspace{0.2cm} if \hspace{0.2cm}and \hspace{0.2cm}only\hspace{0.2cm} if \hspace{0.5cm} \frac{\alpha_1 }{\alpha_2}\le \frac{\xi_2 }{\xi_1}.
\label{equiv}
\eeqa

{\bf Proof.}
Let $\ket{\Gamma_j}$, $j=1,2$ be the superpositions given by Eq. (\ref{superpos}).
 We have $\alpha_1\ge \frac{\eta_1}{1-\xi_1+\eta_1}>\frac{1}{2} $, therefore the largest Schmidt coefficient of $\ket{\Gamma_1}$ is $\alpha_1\xi_1$.

The inequality $\alpha_1\ge \frac{\eta_1}{1-\xi_1+\eta_1}$ is equivalent to $\alpha_1(1-\xi_1)>\eta_1(1-\alpha_1)$. Hence the state $\ket{\Gamma_1}$
\beqa
\ket{\Gamma_1}&=&\sqrt{\alpha_1\xi_1}\ket{00}+\sqrt{\alpha_1(1-\xi_1)}\ket{11}\nonumber\\
&&+\sqrt{(1-\alpha_1)\eta_1}\ket{22}+\sqrt{(1-\alpha_1)(1-\eta_1)}\ket{33}\nonumber
\eeqa
is written with the Schmidt coefficients in the decreasing order:
\beq
\alpha_1\xi_1>\alpha_1(1-\xi_1)>(1-\alpha_1)\eta_1>(1-\alpha_1)(1-\eta_1).
\eeq

By using the condition $\alpha_2>\frac{1}{2}$, we obtain that $\alpha_2\xi_2$ is the largest Schmidt coefficient of the state $\ket{\Gamma_2}$.

Therefore, the conditions of the hypothesis give us a certain order of the Schmidt coefficients of $\ket{\Gamma_1}$ and the largest Schmidt coefficient of $\ket{\Gamma_2}$. We will prove in the following the necessity and sufficiency of the equivalence (\ref{equiv}).

$'\Rightarrow '$ The necessity:
If the transformation $\ket{\Gamma_1}\to \ket{\Gamma_2}$ is possible, then the majorization inequalities (\ref{maj}) are satisfied. The first inequality reads
$\alpha_1\xi_1\le \alpha_2\xi_2,$
which is the conclusion.

$'\Leftarrow '$ The sufficiency:
We know that
\beq
\alpha_1\xi_1\le \alpha_2\xi_2.
\label{N1}
\eeq
 This is the first majorization inequality of the Nielsen's theorem. In addition, we have to prove that the other two inequalities (\ref{maj}) are satisfied. Before proceeding we have to determine the order of the Schmidt coefficients of $\ket{\Gamma_2}$.

Let us observe
\beq
\alpha_2\ge \alpha_1\; \frac{\xi_1}{\xi_2}\ge \frac{\xi_1\eta_1}{1-\xi_1+\eta_1}\; \frac{1}{\xi_2}>\frac{\eta_2}{1-\xi_2+\eta_2},
\label{alfa2}
\eeq
where the last inequality is given by (\ref{ineq1}) and is demonstrated in the Appendix A. We have used the fact that the four parameters satisfy (\ref{inetaxi}). The condition given in the hypothesis $\xi_2 \ge  \frac{\xi_1\eta_1}{1-\xi_1+\eta_1}$ is required in order to have $\alpha_2\le 1$. The inequality (\ref{alfa2}) reads:
$\alpha_2(1-\xi_2)>\eta_2(1-\alpha_2).$
Therefore we have the following ordered Schmidt coefficients of $\ket{\Gamma_2}$:
\beq
\alpha_2\xi_2>\alpha_2(1-\xi_2)>(1-\alpha_2)\eta_2>(1-\alpha_2)(1-\eta_2).
\eeq
Since we have $\alpha_1\le \alpha_2\; \frac{\xi_2}{\xi_1}<\alpha_2$, we get
\beq
\alpha_1\xi_1+\alpha_1(1-\xi_1)<\alpha_2\xi_2+\alpha_2(1-\xi_2).
\label{N2}
\eeq

Now we use the inequality (\ref{ineq2}), which is proved in the Appendix B:
\beq
(1-\alpha_1)(1-\eta_1)>(1-\alpha_2)(1-\eta_2)
\label{ultineg}
\eeq
or equivalently
\beq
\alpha_1\xi_1+\alpha_1(1-\xi_1)+(1-\alpha_1)\eta_1<\alpha_2\xi_2+\alpha_2(1-\xi_2)+(1-\alpha_2)\eta_2.
\label{N3}
\eeq
The inequalities (\ref{N1}), (\ref{N2}), and (\ref{N3}) represent the majorization inequalities (\ref{maj}) required by the Nielsen's theorem, hence the transformation $\ket{\Gamma_1}\to \ket{\Gamma_2}$ can be performed by LOCC.

{\bf Proposition 2.}

{\it Let $\ket{\phi_1}, \ket{\psi_1}, \ket{\phi_2}, \ket{\psi_2}$ be the states given by (\ref{starile}), being characterized by $\xi_j, \eta_j$ satisfying (\ref{inetaxi}). If $\xi_2 \in \bigg[ \frac{1-\eta_1}{2-\xi_1-\eta_1},\frac{\xi_1\eta_1}{1-\xi_1+\eta_1}\bigg)$, then for
$
\alpha_1\in \bigg[ \xi_2,\frac{\eta_1}{1-\xi_1+\eta_1}\bigg)
$
and $\alpha_2>\frac{1}{2}$ we have the following:}
\beqa
&&\sqrt{\alpha_1}\ket{\phi_1}+\sqrt{1-\alpha_1}\ket{\psi_1}\to \sqrt{\alpha_2}\ket{\phi_2}+\sqrt{1-\alpha_2}\ket{\psi_2} \nonumber \\
&&by \hspace{0.2cm} LOCC\hspace{0.2cm}
if \hspace{0.2cm}and \hspace{0.2cm}only\hspace{0.2cm} if \hspace{0.5cm} \frac{\alpha_1 }{\alpha_2}\le \frac{\xi_2 }{\xi_1}.
\eeqa

{\bf Proof.} Since $\alpha_1\in \bigg[ \xi_2,\frac{\eta_1}{1-\xi_1+\eta_1}\bigg)$ with $\xi_2\ge \frac{1-\eta_1}{2-\xi_1-\eta_1} $, we obtain the following ordered Schmidt coefficients of $\ket{\Gamma_1}$:
\beq
\alpha_1\xi_1>(1-\alpha_1)\eta_1>\alpha_1(1-\xi_1)>(1-\alpha_1)(1-\eta_1).
\eeq
By using the condition $\alpha_2>\frac{1}{2}$, we obtain that $\alpha_2\xi_2$ is the largest Schmidt coefficient of the state $\ket{\Gamma_2}$.

$'\Rightarrow '$ The necessity is obvious.

$'\Leftarrow '$ The sufficiency: We use the following inequalities
\beq
\alpha_2\ge \alpha_1\; \frac{\xi_1}{\xi_2}>\alpha_1\ge \xi_2>\frac{\eta_2}{1-\xi_2+\eta_2},
\eeq
where the last inequality is demonstrated in the Appendix A (\ref{ineqxi2}). Hence the ordered Schmidt coefficients of $\ket{\Gamma_2}$ are: $\alpha_2\xi_2>\alpha_2(1-\xi_2)>(1-\alpha_2)\eta_2>(1-\alpha_2)(1-\eta_2).$
The first majorization inequality is verified. Further we start with $\xi_1(1-\xi_2)>\eta_1(1-\xi_2)$ which can be written as
\beq
\frac{\xi_1}{\xi_2}>\xi_1-\eta_1+\frac{\eta_1}{\xi_2}.
\eeq
By using the condition $\alpha_1\ge \xi_2$, we obtain
\beq
\frac{\alpha_2}{\alpha_1}\ge \frac{\xi_1}{\xi_2}>\xi_1-\eta_1+\frac{\eta_1}{\alpha_1}.
\eeq
This is equivalent to
\beq
\alpha_2=\alpha_2\xi_2+\alpha_2(1-\xi_2)>\alpha_1\xi_1+\eta_1(1-\alpha_1).
\label{in2P2}
\eeq
With the help of the inequality (\ref{ineq2}) given in the Appendix B, we find that the third majorization inequality is satisfied:
\beq
\alpha_1\xi_1+(1-\alpha_1)\eta_1+\alpha_1(1-\xi_1)<\alpha_2\xi_2+\alpha_2(1-\xi_2)+(1-\alpha_2)\eta_2.
\label{in3P2}
\eeq
The inequalities $\alpha_1\xi_1\le \alpha_2\xi_2$, (\ref{in2P2}), and (\ref{in3P2}) are the majorization inequalities and this leads to the fact that the conversion can be realized by LOCC.

{\bf Proposition 3.}

{\it Let $\ket{\phi_1}, \ket{\psi_1}, \ket{\phi_2}, \ket{\psi_2}$ be the states given by (\ref{starile}), being characterized by $\xi_j, \eta_j$ satisfying (\ref{inetaxi}). If $\xi_2 \in \bigg( \frac{1}{2},\frac{1-\eta_1}{2-\xi_1-\eta_1}\bigg)$, then for $\alpha_1\in \bigg[ \xi_2, \frac{1-\eta_1}{2-\xi_1-\eta_1}\bigg]$
and $\alpha_2>\frac{1}{2}$ we have the following:}
\beqa
&&\sqrt{\alpha_1}\ket{\phi_1}+\sqrt{1-\alpha_1}\ket{\psi_1}\to \sqrt{\alpha_2}\ket{\phi_2}+\sqrt{1-\alpha_2}\ket{\psi_2} \nonumber \\
&&by \hspace{0.2cm} LOCC\hspace{0.2cm}
if \hspace{0.2cm}and \hspace{0.2cm}only\hspace{0.2cm} if \hspace{0.5cm} \frac{\alpha_1 }{\alpha_2}\le \frac{\xi_2 }{\xi_1}.
\eeqa

{\bf Proof.} Since $\alpha_1\in \bigg[ \xi_2,\frac{1-\eta_1}{2-\xi_1-\eta_1}\bigg)$ with $\xi_2<\frac{1-\eta_1}{2-\xi_1-\eta_1}$, we have:
\beq
\alpha_1\xi_1>(1-\alpha_1)\eta_1>(1-\alpha_1)(1-\eta_1)>\alpha_1(1-\xi_1).
\eeq
By using the condition $\alpha_2>\frac{1}{2}$, we obtain that $\alpha_2\xi_2$ is the largest Schmidt coefficient of the state $\ket{\Gamma_2}$.

$'\Rightarrow '$ The necessity is obvious.

$'\Leftarrow '$ The sufficiency: We have
$\alpha_2>\alpha_1\ge \xi_2>\frac{\eta_2}{1-\xi_2+\eta_2}$,
where the last inequality is demonstrated in the Appendix A (\ref{ineqxi2}). Hence the ordered Schmidt coefficients of $\ket{\Gamma_2}$ are:
$\alpha_2\xi_2>\alpha_2(1-\xi_2)>(1-\alpha_2)\eta_2>(1-\alpha_2)(1-\eta_2).$
The first majorization inequality is verified. Then we have $\alpha_2\xi_2+\alpha_2(1-\xi_2)>\alpha_1\xi_1+\eta_1(1-\alpha_1) $ and this is
the second majorization inequality.

Further we start from the inequality $\xi_2\; \frac{1-\xi_1}{1-\eta_2}>1-\xi_1$ and due to the fact that $\alpha_1\ge \xi_2$ we obtain
$\frac{\xi_1}{\xi_2}>\frac{1}{\alpha_1}-\frac{1-\xi_1}{1-\eta_2}.$
On the other hand we have
\beq
\frac{\alpha_2}{\alpha_1}\ge \frac{\xi_1}{\xi_2}>\frac{1}{\alpha_1}-\frac{1-\xi_1}{1-\eta_2},
\eeq
which leads to
$\alpha_1(1-\xi_1)>(1-\eta_2)(1-\alpha_2).$
This inequality is equivalent to
\beqa
\alpha_1\xi_1&+&(1-\alpha_1)\eta_1+(1-\eta_1)(1-\alpha_1)<\alpha_2\xi_2\nonumber\\
&+&\alpha_2(1-\xi_2)+\eta_2(1-\alpha_2)
\label{in3P3}
\eeqa
and represents the third majorization inequality. Hence the local transformation is possible.

\subsection{An example}
Let us apply our result for performing the transformation between the states defined at the beginning of Section 2, namely the states of Eqs. (\ref{initial-ex}) and (\ref{final-ex}): $\xi_1=\frac{9}{10}$, $\eta_1=\frac{4}{5}$, $\xi_2=\frac{7}{10}$, and $\eta_2=\frac{3}{5}$.
Firstly we compute
$$
\frac{\xi_1\eta_1}{1-\xi_1+\eta_1}=0.8 \hspace{0.5cm} \mbox{and} \hspace{0.5cm}
\frac{1-\eta_1}{2-\xi_1-\eta_1}=0.67.
$$
We see that $\xi_2\in \bigg( \frac{1-\eta_1}{2-\xi_1-\eta_1},\frac{\xi_1\eta_1}{1-\xi_1+\eta_1}\bigg)$, therefore we apply the Proposition 2. We have $\frac{\eta_1}{1-\xi_1+\eta_1}=0.89$, which means that $\alpha_1\in [0.7,0.89)$. Let us take $\alpha_1=\frac{3}{4}$. We must have
\beq
\alpha_2\ge \alpha_1\; \frac{\xi_1}{\xi_2}=0.964.
\eeq
By defining $\alpha_2=0.98$ we know that the conversion is possible by LOCC. Indeed one can verify that the ordered Schmidt coefficients of the two superpositions are:
\beqa
\ket{\Gamma_1}:&&\hspace{0.5cm}\frac{675}{1000}; \hspace{0.3cm}\frac{200}{1000}; \hspace{0.3cm}\frac{75}{1000}; \hspace{0.3cm} \frac{50}{1000};\nonumber\\
\ket{\Gamma_2}:&&\hspace{0.5cm}\frac{686}{1000}; \hspace{0.3cm}\frac{294}{1000}; \hspace{0.3cm}\frac{12}{1000}; \hspace{0.3cm} \frac{8}{1000}
\eeqa
and that they satisfy the majorization inequalities.

\section{Conclusions}

In the present Letter, we have derived three propositions which represent the necessary and sufficient conditions to enable the transformations of superpositions of entangled states by LOCC. By applying the Nielsen's theorem we have shown that the two coefficients of superpositions $\alpha_1$ and $\alpha_2$ depend on some inequalities which involve the Schmidt coefficients of only three states $\ket{\phi_1}$, $\ket{\psi_1},$ and $\ket{\phi_2}$.
The analysis reported in this Letter could lead to a deeper understanding of the behavior of entanglement under LOCC and may be relevant in the future work on entanglement manipulations.

\section*{Acknowledgement}
This work was supported by the Romanian Ministry of Education and Research through Grant IDEI-995/2007 for the University of Bucharest.

\begin{appendix}
\renewcommand\theequation{\thesection.\arabic{equation}}
\setcounter{equation}{0}

\section{The proof of the inequality (\ref{alfa2})}
In this appendix we will prove the following inequality, which is used for proving Proposition 1. If
$\frac{1}{2}<\eta_2<\xi_2<\eta_1<\xi_1<1,  $
then
\beq
\frac{\xi_1\eta_1}{1-\xi_1+\eta_1}>\frac{\xi_2\eta_2}{1-\xi_2+\eta_2}
\label{ineq1}
\eeq

{\it Proof.} Let us observe that
$
(\xi_2-\eta_2)(1-\xi_2)>0,
$
which is equivalent to
\beq
\xi_2>\frac{\eta_2}{1-\xi_2+\eta_2}.
\label{ineqxi2}
\eeq
We have
$
\eta_1^2>\xi_2^2>\frac{\xi_2\eta_2}{1-\xi_2+\eta_2}
$
from which we obtain
\beq
\eta_1>\frac{\xi_2\eta_2(1+\eta_1)}{\eta_1(1-\xi_2+\eta_2)+\xi_2\eta_2}.
\label{intermed}
\eeq
On the other hand we know that $\xi_1>\eta_1$, which together with Eq. (\ref{intermed}) leads to:
\beq
\xi_1\eta_1(1-\xi_2+\eta_2)>\xi_2\eta_2(1-\xi_1+\eta_1).
\eeq

\section{The proof of the inequality (\ref{ultineg})}

\setcounter{equation}{0}

Here we will prove a second inequality, namely: If
$\frac{1}{2}<\eta_2<\xi_2<\eta_1<\xi_1<1, $ and
$\alpha_1\xi_1\le \alpha_2\xi_2, $
then the following inequality holds:
\beq
(1-\alpha_1)(1-\eta_1)>(1-\alpha_2)(1-\eta_2).
\label{ineq2}
\eeq

{\it Proof.} The inequality
$
\alpha_1(1-\eta_1)(\xi_1-\xi_2)>0
$
can be written as follows
$
(1-\eta_1)(\alpha_1\xi_1-\xi_2)+(1-\eta_1)\xi_2(1-\alpha_1)>0.
$
Further we get
\beq
(1-\eta_2)(\alpha_1\xi_1-\xi_2)+(1-\eta_1)\xi_2(1-\alpha_1)>0.
\eeq
This last inequality is equivalent to
\beq
\alpha_1\; \frac{\xi_1}{\xi_2}>\frac{\alpha_1(1-\eta_1)+\eta_1-\eta_2}{1-\eta_2}.
\eeq
From the hypothesis we have $\alpha_2\ge \alpha_1\; \frac{\xi_1}{\xi_2}$, therefore we obtain
\beq
\alpha_2>\frac{\alpha_1(1-\eta_1)+\eta_1-\eta_2}{1-\eta_2},
\eeq
which is equivalent to
$
(1-\alpha_1)(1-\eta_1)>(1-\alpha_2)(1-\eta_2).
$
\end{appendix}



\end{document}